\documentclass{ws-procs9x6}

\usepackage{latexsym,amsopn,mathrsfs,bm,mathbbol}

\newcommand{\op}[1]{\ensuremath{\bm{\mathrm{#1}}}}
\newcommand{\funcint}[2]{\ensuremath{\int_{\mathcal{C}} \mathrm{d}
    (#2) #1}}
\newcommand{\feynint}[1]{\ensuremath{\int \frac{\mathrm{d}^{d} {#1}}{(2
\pi)^{d}}}}
\newcommand{\dd}{\ensuremath{\mathrm{d}}}
\newcommand{\fint}[1]{\ensuremath{\int \frac{\mathrm{d}^4 #1}{(2\pi)^4}}}
\DeclareMathOperator{\Tr}{Tr}
\DeclareMathOperator{\im}{Im}

\newcommand{\Z}{\ensuremath{\mathbb{Z}}}
\newcommand{\ii}{\ensuremath{\mathrm{i}}}
\newcommand{\Gvac}{G^{(\mathrm{vac})}}
\newcommand{\Gmat}{G^{(\mathrm{mat})}}
\newcommand{\funcd}[2]{\frac{\delta #1}{\delta #2}}

\usepackage{makeidx}
\makeindex

\begin{document}

\title{Renormalization of self-consistent $\Phi$-derivable
  approximations\footnote{Based on a talk presented at the conference
    ``Progress in Nonequilibrium Greens Functions, Dresden, Germany,
    19.-22. August 2002''}}

\author{H. van Hees}
\address{Fakult{\"a}t f{\"u}r Physik \\
  Universit{\"a}t Bielefeld \\
  Universit{\"a}tsstra{\ss}e 25\\
  D-33615 Bielefeld \\
 E-mail: hees@physik.uni-bielefeld.de}

\author{J. Knoll}
\address{Gesellschaft f{\"u}r Schwerionenforschung\\
  Planckstra{\ss}e 1\\
  D-64291 Darmstadt\\
  E-Mail: j.knoll@gsi.de}


\maketitle

\abstracts{Within finite temperature field theory, we show that truncated
  non-perturbative self-consistent Dyson resummation schemes\index{Dyson
    resummation schemes} can be renormalized with local vacuum
  counterterms. For this the theory has to be renormalizable in the usual
  sense and the self-consistent scheme must follow Baym's
  $\Phi$-derivable\index{Phi-derivable approximation} concept. Our
  BPHZ-renormalization scheme leads to renormalized self-consistent
  equations of motion. At the same time the corresponding 2PI-generating
  functional and the thermodynamic potential\index{thermodynamic potential}
  can be renormalized with the same counterterms used for the equations of
  motion.  This guarantees the standard $\Phi$-derivable properties like
  thermodynamic consistency and exact conservation laws also for the
  renormalized approximation schemes.  We give also a short overview over
  symmetry properties of the various functions defined within the 2PI
  scheme for the case that the underlying classical field theory has a
  global linearly realized symmetry.}

\section{Introduction} 

Describing hot and dense systems of strongly interacting particles, one is
led to the use of \emph{dressed propagators} within non-perturbative
Dyson-resummation schemes. Especially this becomes unavoidable if
\emph{damping width effects}\index{damping width} are significant for the
physical situation in question.

Based on functional formulations by Luttinger and Ward\cite{lw60} and Lee
and Yang\cite{leeyang61} Baym and Kadanoff\cite{bk61} studied a special
class of self-consistent Dyson approximations, which later was reformulated
in terms of a variational principle, defining the so-called
\emph{$\Phi$-derivable approximations}\cite{baym62}. The variational
principle, applied to approximations of the $\Phi$-functional, leads to
closed coupled equations of motion for the mean field and the propagator,
which guarantee the exact conservation of the expectation values of
conserved currents and thermodynamical consistency, since at the same time
the approximated $\Phi$-functional is an approximation of the thermodynamic
potential.

Later this concept was generalized to the relativistic case and rederived
within the path integral\index{path integral} formalism by Cornwall,
Jackiw, and Tomboulis\cite{cjt74}. It is no formal problem to extend this
formulation to the general Schwinger-Keldysh real-time
contour\cite{Sch61,kel64} and thus to generalize the concept to
non-equilibrium problems.

Here we discuss the problem, how to \emph{renormalize} the equations of
motion, derived from $\Phi$-derivable approximations for relativistic
quantum field theories. Generalizing the work of Bielajew and
Serot\cite{biel83,bielserot84} we show that any $\Phi$-derivable
approximation of a perturbatively renormalizable theory is also
renormalizable in the usual sense. Further we prove that at finite
temperature only \emph{temperature-independent
  counterterms}\index{counterterm} are necessary to give finite equations
of motion\cite{vHK2001-Ren-I}. The counterterms can be interpreted as
renormalization of the wave function and the vacuum parameters of the
quantum field theory, like the particle masses and the coupling constants.

Further we demonstrate the possibility to treat numerically
$\Phi$-derivable approximations with generic two-point contributions to the
self-energy beyond pure gap-equation approximations with ``tadpole
self-energies'') giving rise to a finite in-medium damping width of the
involved particles\cite{vHK2001-Ren-II}.

Another important question is whether the approximations respect underlying
symmetries\index{Symmetries} of the classical action functional. Contrary
to perturbation theory in general the solution of the $\Phi$-derivable
equations of motion violates the Ward-Takahashi identities of symmetries.
This was discussed first by Baym and Grinstein\cite{baymgrin} on the
example of the O($N$)-symmetric linear $\sigma$-model. The reason can be
traced back to a violation of \emph{crossing symmetry}\index{Crossing
  symmetry} for approximations of the $\Phi$-functional: The solution of
the $\Phi$-derivable equations of motion is equivalent to a resummation of
the self-energy in any order of the expansion parameter of the
$\Phi$-functional approximation. This involves intrinsically the
resummation of higher vertex functions, which is incomplete, because
certain channels are missing, being taken into account only by
approximations of the $\Phi$ functional at higher orders.

We show that this can partially be cured by defining a non-perturbative
approximation to the usual effective quantum action. This admits one to
define vertex-functions which fulfill the Ward-Takahashi identities of the
symmetry. These crossing symmetric vertex-functions are defined by
equations of motion which solution is equivalent to a further resummation
of the channels, missing intrinsically in the $\Phi$-derivable self-energy
resummation\cite{vHK2001-Ren-III}.

\section{The 2PI generating functional\index{2PI functional}}
\label{sect:2pi}

\begin{figure}[th]
\epsfxsize 7cm
\centerline{\epsfbox{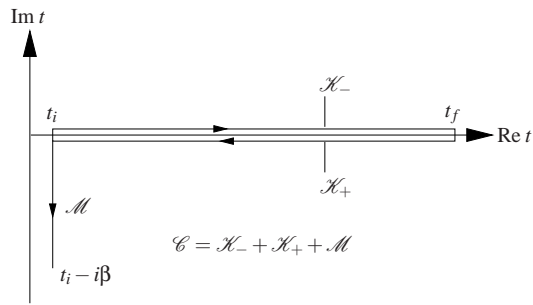}}
\caption{\label{fig:1}The Schwinger-Keldysh closed time path modified for the
application to thermal equilibrium quantum field theory.} 
\end{figure}

We start with the defining path integral for the two-particle (2PI)
irreducible quantum action for the state of thermal
equilibrium. For this case we use the Schwinger-Keldysh closed real-time
path extended by an imaginary part making use of the fact that the
unnormalized thermal density operator $\exp(-\beta \op{H})$, with $\beta$
denoting the inverse temperature of the system and $\op{H}$ its Hamilton
operator, can be included within the path integral as a time evolution
parallel to the imaginary axis (see Fig. \ref{fig:1}).
We consider the local relativistic renormalizable quantum field theory for
one scalar field $\phi$ with the dynamics defined
by the classical action
\begin{equation}
\label{1}
S[\phi]=\funcint{\left [ \frac{1}{2} (\partial_{\mu} \phi_1)
  (\partial^{\mu} \phi_1) - 
  \frac{m^2}{2} \phi_1^2 -\frac{\lambda}{4!} \phi_1^4 \right]}{1}.
\end{equation}
Here and in the following $\funcint{f_{123\ldots}}{123\ldots}$ denotes an
integral over a function $f$ of space-time arguments $x_1,x_2,\ldots$. The
time variable is assumed to be defined along the contour depicted in Fig.
\ref{fig:1}.

The generating functional is given by
\begin{equation}
 \label{8}
 \Gamma[\varphi,G] =S[\varphi]+\frac{\ii}{2} \Tr(M^2 G^{-1}) + \frac{\ii}{2}
 \funcint{D_{12}^{-1}(G_{12}-D_{12})}{12} + \Phi[\varphi,G]
\end{equation}
with 
\begin{equation}
\label{9}
D^{-1}_{12}=\frac{\delta^2 S[\varphi]}{\delta \varphi_1 \delta \varphi_2}.
\end{equation}
In terms of diagrams the functional $\Phi$ consists of closed two-loop
diagrams, which are built with lines representing \emph{exact
  propagators}\index{exact propagators} $G$ and point-vertices with respect
to the field $\phi$ derived from the action $S[\varphi+\phi]$.

The equations of motion are determined by the stationary point of the
functional (\ref{8}):
\begin{equation}
\label{10}
\funcd{\Gamma}{\varphi_1} \stackrel{!}{=}0, \quad
\funcd{\Gamma}{G_{12}} \stackrel{!}{=}0.  
\end{equation}
Using (\ref{8}) these equations of motion read
\begin{eqnarray}
\funcd{S}{\varphi_1} &=& -\frac{\ii}{2} \funcint{
\funcd{D_{1'2'}^{-1}}{\varphi_1} G_{1'2'}}{1'2'} - \funcd{\Phi}{\varphi_1},
\label{11} \\
\Sigma_{12} &:=& D_{12}^{-1} - G_{12}^{-1} = 2\ii \funcd{\Phi}{G_{12}} \label{12}.
\end{eqnarray}
From the latter equation we see that the derivative of $\Phi$ with respect
to $G$ gives the \emph{exact self-energy} of the theory at presence of a
mean field $\varphi$, which in turn is determined from (\ref{11})
self-consistently.  Since lines in diagrams contributing to an expansion of
$\Phi$ stand for exact Green's functions $G$ all these lines must not
contain any self-energy insertions, i.e., the self-energy is represented as
the sum of all \emph{skeleton diagrams}\index{skeleton diagram}. Since the
derivative of a diagram with respect to $G$ means to open any line
contained in it and then adding all the so obtained diagrams, the
functional $\Phi$ consists of all \emph{closed two-particle irreducible
  diagrams} with at least two loops.

A $\Phi$-derivable \emph{approximation} is defined as the truncation of the
functional $\Phi$ to a finite (e.g., coupling-constant or
$\hbar$-expansion) or an explicitly resummable infinite subset of 2PI
diagrams. The mean field $\varphi$ and Green's function $G$ are then
determined by the self-consistent closed equations of motion
(\ref{11}-\ref{12}).

\section{Renormalization of $\phi^4$-theory}
\label{sect:renorm}

For the renormalization of self-consistent approximation schemes we use the
Bogoliubov-Parasiuk-Hepp-Zimmermann (BPHZ) renormalization\index{BPHZ
  renormalization} description. As an example we treat $\phi^4$-theory in
the phase of unbroken $\Z_2$-symmetry, i.e., we set $m^2>0$. Then as well
at zero as at finite temperature we have $\varphi=0$ as the unique solution
of the equations of motion. Only the self-consistent propagator has to be
determined.

\subsection{Renormalization at $T=0$}
\label{subsect:renorm-vac}

At $T=0$ the only difference to the perturbative treatment of the
renormalization problem is that we have to apply it to diagrams with lines
standing for self-consistent propagators instead for free ones. It is also
clear that we can restrict ourselves to the $\{--\}$-part (i.e., the time
ordered part) of the real-time contour since in the vacuum case the
time-ordered Green's function is identical to the retarded (advanced) one
for positive (negative) $p^0$-components.

The BPHZ renormalization description rests solely on Weinberg's
power-counting theorem which is independent of the the special form of
propagators. Thus we only have to show that the self-consistent propagators
of $\Phi$-derivable approximations belong to the class of functions with
the asymptotic behavior $O[(l^2)^{-1} (\ln l^2)^{\beta}]$ for large momenta
$l^2$, where $\beta$ is a constant. Assuming that this is the case
Weinberg's theorem tells us that a connected truncated diagram $\gamma$
with $E$ external lines has a superficial degree of divergence
$\delta(\gamma)=4-E$. So due to the $\Phi$-derivable equations of motion
the self-consistent self-energy shows an asymptotic behavior like $O[p^2
(\ln p^2)^{\beta'}]$. So starting an iteration for the self-consistent
propagator with the perturbative propagator (and provided this iteration
converges) we can conclude that indeed the propagator is of the usual
asymptotic behavior.

The BPHZ renormalization technique aims at the construction of the
\emph{integrand} of the renormalized integral without using an intermediate
step of regularization. If a diagram is divergent without proper divergent
subdiagrams it is sufficient to subtract the Taylor expansion of the
integrand with respect to the external momenta up to the order given by the
superficial degree of divergence which is in our case $4-E$, with $E$
denoting the number of external legs. 

For renormalization theory it is crucial that the same holds true for
diagrams which contain divergences from proper subdiagrams, if the
according subdivergences are subtracted first, even if it contains
\emph{overlapping} divergences\index{overlapping divergences} and thus
that one needs only \emph{local counterterms}\index{local counterterms} to
the quantum action which have the same form as that of the classical action
but with the infinities lumped into the ``bare parameters'' rather than the
physical ones\cite{bp57,Zim69}.

The described BPHZ-scheme chooses the renormalization point for divergent
diagrams at external momenta set to $0$. It is clear that by another finite
renormalization of the same diagrams we can switch to any renormalization
scheme appropriate for the application under consideration. In our case of
$\phi^4$-theory we choose the on-shell renormalization
scheme\index{on-shell-renormalization scheme}, which defines the mass
parameter $m$ to denote the physical mass of the particles. We use the
following on-shell renormalization conditions:
\begin{eqnarray}
\Sigma^{(\mathrm{vac})}(p^2=m^2) &=& 0, \quad
 \partial_{p^2} \Sigma^{(\mathrm{vac})}(p^2=m^2) =0, \label{19}\\
\Gamma^{(4,\mathrm{vac})}(s,t,u=0) &=& \frac{\lambda}{2}. \label{20}
\end{eqnarray}
Here $s$, $t$, $u$ are the usual Mandelstam variables for two-particle
scattering, $p$ is the external momentum of the self-energy and $m^2$ is
the (renormalized) mass of the particles due to the renormalization
conditions (\ref{19}). The second condition defines the wave-function
normalization such that the residuum of the propagator at $p^2=m^2$ is equal to
unity.

\subsection{Numerical calculation}
\label{subsect:renorm-num}

To illustrate the abstract considerations of the previous section we show
how to solve the self-consistent $\Phi$-derivable Dyson equation for the
$\phi^4$ model in the case of unbroken $\Z_2$-symmetry, i.e., for
$\varphi=0$. We take into account the $\Phi$-functional up to three-loop
order:
\begin{equation}
\label{21}
\Phi=\parbox[c][10mm][c]{16mm}{\includegraphics{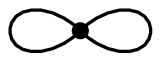}} +
\frac{1}{2}
\parbox[c][15mm][c]{15mm}{\includegraphics{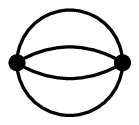}}
\end{equation} 
For the self-energy we find from Eq. (\ref{12})
\begin{equation}
\label{22}
-\ii
\Sigma=\;\raisebox{4.5mm}{\parbox{7mm}{\centerline{\includegraphics{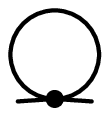}}}}
\quad + \;\parbox{20mm}{\centerline{\includegraphics{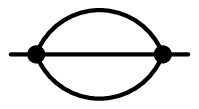}}}
\end{equation}
The main numerical problem is that it is of course not possible to
integrate directly the renormalized integrands of the self-energy diagrams
depicted in Eq. (\ref{22}) because of the on-shell poles of the
propagator. Instead we use its Lehmann spectral
representation\index{spectral representation} 
\begin{equation}
\label{23}
\Gvac(p^2)=\int_0^{\infty} \frac{\dd(m^2)}{\pi} \frac{\im
  \Gvac(m^2)}{m^2-p^2-\ii \eta},
\end{equation}
where $\eta$ denotes a small positive number to be taken to $0^+$ in the
sense of a weak limit after performing the loop integrals.

First we calculate the one-loop function
\begin{equation}
\label{24}
L^{(\mathrm{reg})}(q^{2})=\ii \,
\raisebox{0.2mm}{\parbox{17mm}{\centerline{\includegraphics{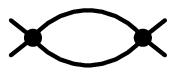}}}}=\ii
\feynint{l} \Gvac[(l+q)^{2}] 
\Gvac(l^{2}) 
\end{equation}
which appears as a subdiagram contained in the ``sunset diagram'' in Eq.
(\ref{22}). Here and in the following we use \emph{dimensional
  regularization}\index{dimensional regularization} to give the
un-renormalized integrals a definite meaning.  At the end of the BPHZ
subtraction procedure we can let $d=4$. The loop function (\ref{24}) is
logarithmically divergent and has to be subtracted such that
$L^{(\mathrm{ren})}(q^2=0)=0$ due to the renormalization conditions
(\ref{19}).

Using the spectral representation (\ref{23}) the \emph{renormalized} loop
function can be expressed with help of a kernel $K_1^{(\mathrm{ren})}$:
\begin{equation}
\label{25}
\begin{array}{ll}
\displaystyle L^{\mathrm{ren}}(q^2)=  \int_0^{\infty}
\frac{\dd m_1^2}{\pi} \int_0^{\infty} 
\displaystyle \frac{\dd m_2^2}{\pi}  & \displaystyle  K_1^{(\mathrm{ren})}(q^2,m_1^2,m_2^2)
\\ & \displaystyle \times \im 
\Gvac(m_1^2) \im \Gvac(m_2^2). 
\end{array}
\end{equation}
The renormalized kernel can be calculated analytically with help of
standard formulae of perturbation theory (for details
see\cite{vHK2001-Ren-II}).

Due to the renormalization conditions the tadpole contribution to the
self-energy is canceled. For the remaining sunset-diagram we can use the
dispersion relation for $L^{(\mathrm{ren})}$ and $\Gvac$ to define a kernel
$K_2$ such that the renormalized self-energy reads
\begin{equation}
\label{28}
\begin{array}{ll}
\displaystyle
\Sigma^{(\mathrm{vac})}(p^2) = \int_{4m^2}^{\infty} \frac{\dd m_3^2}{\pi}
\int_{0}^{\infty} \frac{\dd m_4^2}{\pi}& \displaystyle K_2^{\mathrm{(ren)}}
(p^2,m_3^2,m_4^2) \\[3mm] 
\displaystyle
& \displaystyle \times \im L^{(\mathrm{ren})}(m_3^2) \im \Gvac(m_4^2).
\end{array}
\end{equation}
For the numerical calculation one has to take into account that $\im \Gvac$
contains the pole contribution $\propto \delta(p^2-m^2)$ which has to be
treated explicitely in both formulae (\ref{25}) and (\ref{28}). The
remaining integrals over the $m_k^2$ are relatively smooth finite integrals
which can be done with help of a simple adaptive integration algorithm. We
used an adaptive Simpson algorithm to solve the equations (\ref{25}) and
(\ref{28}) iteratively.

\begin{figure}
\begin{minipage}{0.49\textwidth}
  \centering{\includegraphics[width=\textwidth]{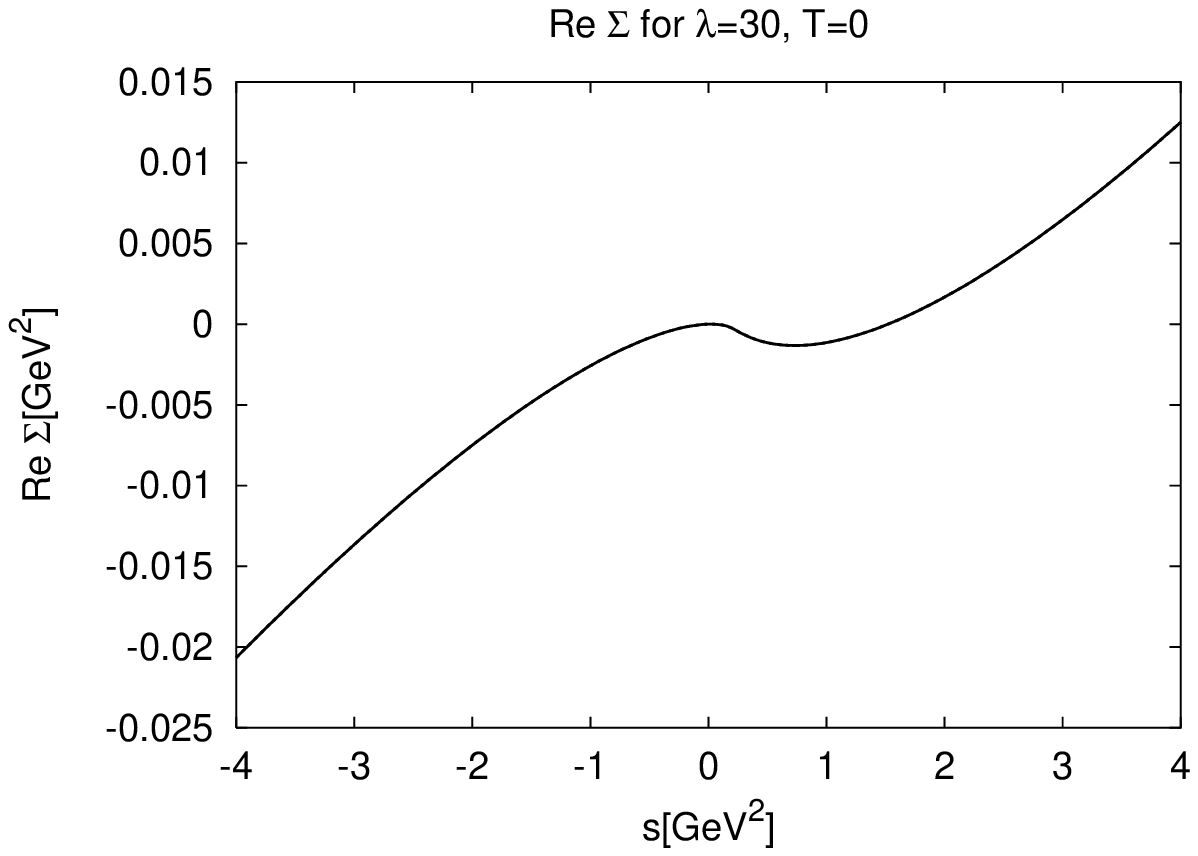}}
\end{minipage}
\begin{minipage}{0.49\textwidth}
\hspace*{2mm}
\centering{\includegraphics[width=\textwidth]{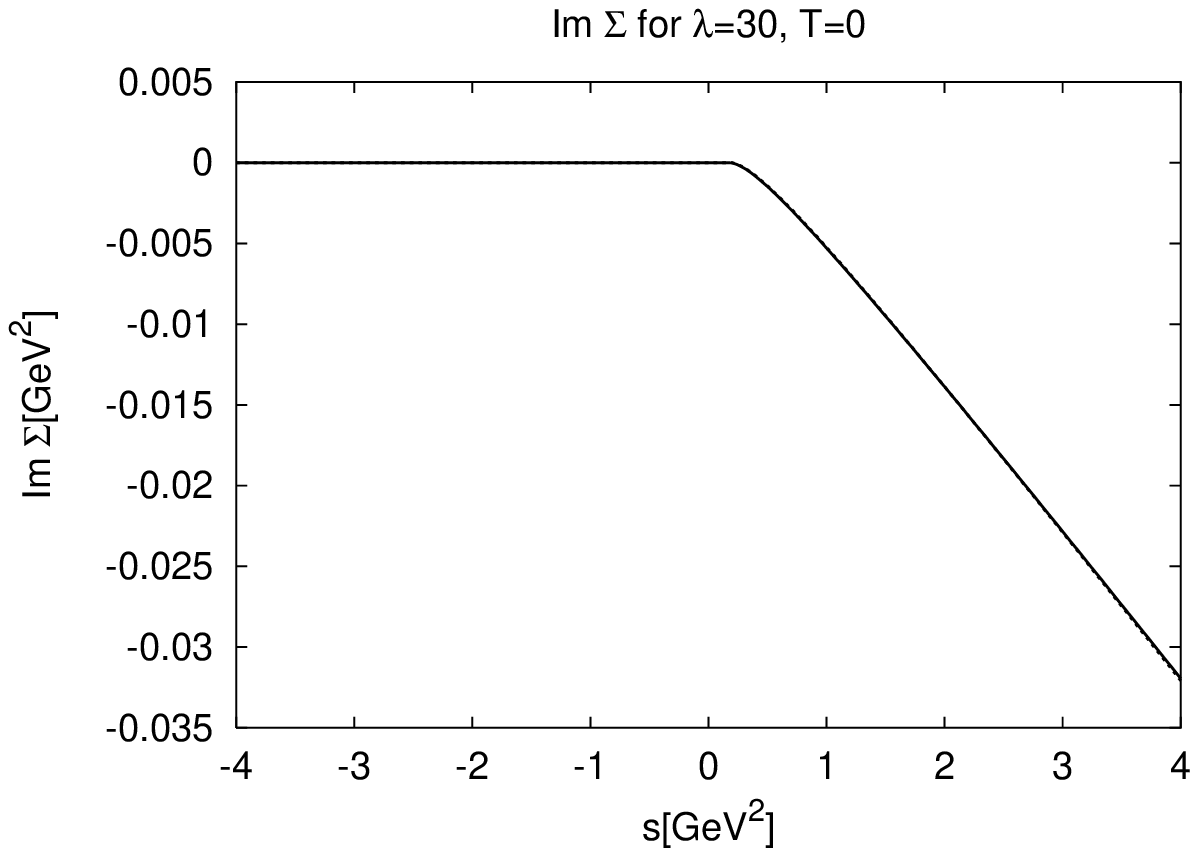}}
\end{minipage}
\caption{Real (left) and imaginary part (right) of the sunset self-
  energy. The perturbative and the self-consistent result lie on top
  of each other due to the large threshold at $s=9 m^2$.}
\label{fig:2}
\end{figure}

As turns out for the vacuum case the main contribution comes from the pole
terms such that even for high coupling constants the perturbative and
self-consistent result lie on top of each other (see Fig. \ref{fig:2}). The
reason is that in our on-shell scheme the threshold for the imaginary part
of the self-energy is at $9m^2$.

\subsection{Renormalization at $T>0$}
\label{sect:ren-fin-T}

Now we show that the renormalization at $T>0$ can be done with the
\emph{same temperature independent vacuum counterterms} as were necessary
to render the vacuum proper vertex functions finite. Thus in complete
analogy to the well-known result of perturbative finite-temperature
renormalization theory the renormalized theory is completely defined at
$T=0$. There is no ambiguity for ``in-medium modifications'' of coupling
constants from renormalization.

We expand the finite-temperature self-energy around the self-consistent
solution of the same $\Phi$-derivable approximation at $T=0$:
\begin{equation}
\label{30}
\Sigma_{12}=\Sigma_{12}^{(\mathrm{vac})}+\Sigma_{12}^{(0)}+\Sigma_{12}^{(r)}.
\end{equation}
Here $\Sigma_{12}^{(\mathrm{vac})}$ is the renormalized vacuum self-energy
calculated in the previous section. The second and third terms in Eq.
(\ref{30}) contain the in-matter parts of the self-energy. Thereby
$\Sigma^{(0)}$ is the part of the self-energy which arises as the
\emph{linear part} from a functional power expansion with respect to the
Green's function around the vacuum Green's function:
\begin{equation}
\label{31}
-\ii \Sigma_{12}^{(0)}= -\ii \funcint{\left (
 \left. \funcd{\Sigma_{12}}{G_{1'2'}} \right|_{T=0} \Gmat_{1'2'}
 \right)}{1'2'} = \parbox{25mm}{\includegraphics{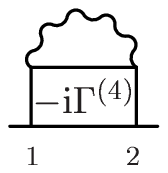}} 
\end{equation}
The wavy line stands for the ``matter part'' of the Green's function
$\Gmat=G-\Gvac$. As we shall see below herein we have to understand only
the \emph{diagonal part} of the vacuum propagator within the momentum-space
matrix formalism. The four-point kernel $\Gamma^{(4)}$ is a four-point
function represented by a particular set of self-energy subdiagrams
consisting of pure vacuum lines, defined by
\begin{equation}
\label{32}
-\ii \Gamma_{12,1'2'}^{(4)}=-\left . \funcd{\Sigma_{12}}{G_{1'2'}}
 \right|_{T=0} = -2 \ii \left . \frac{\delta^2 \Phi}{\delta G_{12} \delta
 G_{1'2'}} \right|_{T=0}.
\end{equation}
Its ``diagonal part'', i.e., with all time arguments placed on one side of
the real-time contour, defines a vacuum renormalization part which is of
superficial degree of divergence $0$. Thus we can conclude that the
\emph{diagonal part} of $\Gmat$ is of momentum power $-4$, so that closing
$\Gamma^{(4)}$ with a wavy $\Gmat$-line yields another logarithmic
divergence, even when the pure vacuum part $\Gamma^{(4)}$ is renormalized.
On the other hand the off-diagonal parts of $\Gmat$ contain
$\theta$-functions and Bose-Einstein-distribution factors which lead to
convergent temperature dependent integrals which we are not allowed to
subtract. Thus from Weinberg's power-counting theorem we can conclude that
the divergent part of $\Sigma^{(0)}$, in the following called
$\Sigma^{(0,\mathrm{div})}$, accounts for all terms of momentum power $0$
and consequently $\Sigma^{(r)}$ is of divergence degree $-2$ and thus
finite after subtracting vacuum subdivergences.

So we are left with the task to renormalize the last loop integral from
closing the $\Gamma^{(4)}$-diagram with a $\Gmat$-line. For this purpose
due to our discussion above we have to split the full propagator as follows
\begin{equation}
\label{33}
\begin{array}{ll}
\displaystyle \ii G_{12} &= \displaystyle
\ii \Gvac_{12} + \ii \funcint{\Gvac_{11'} \Sigma_{1'2'}^{(0,\mathrm{div})}
  G_{2'2}^{(\mathrm{vac})}}{1'2'} + \ii G_{12}^{(r)} \\
&= \parbox{17mm}{\centerline{\includegraphics{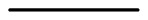}}} +
  \parbox[b]{38mm}{\raisebox{0.3mm}{\centerline{\includegraphics{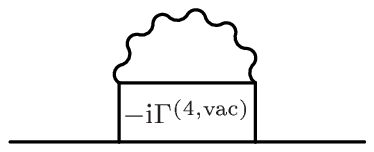}}}} +
  \parbox{17mm}{\includegraphics{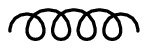}} 
\end{array}
\end{equation}
Using Eq. (\ref{33}) shows that $\Sigma^{(0,\mathrm{div})}$, represented by
the second diagram in (\ref{33}), fulfills the equation of motion
\begin{equation}
\label{34}
\Sigma_{12}^{(0,\mathrm{div})} = \funcint{\Gamma_{12,1'2'}^{(4,\mathrm{vac})}}{1'2'}
 \left ( \funcint{\Gvac_{1'1''} \Sigma_{1''2''}^{(0,\mathrm{div})}
 \Gvac_{2''2'}}{1''2''} + G_{1'2'}^{(r)} \right )
\end{equation}
which is \emph{linear} in $G^{(r)}$. Thus Eq. (\ref{34}) is solved by the
ansatz
\begin{equation}
\label{35}
-\ii \Sigma_{12}^{(0,\mathrm{div})} = \funcint{\Lambda_{12,1'2'}
 G_{1'2'}^{(r)}}{1'2'} = \raisebox{3mm}{\parbox{2cm}{\includegraphics{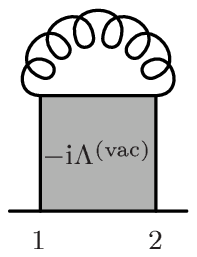}}},
\end{equation}
where the vacuum four-point function $\Lambda^{(\mathrm{vac})}$ fulfills
the \emph{Bethe-Salpeter equation}\index{Bethe-Salpeter equation}
\begin{equation}
\label{36}
\Lambda_{12,1'2'}^{(\mathrm{vac})} = \Gamma_{12,1'2'}^{(4,\mathrm{vac})} +
\ii \funcint{\Gamma_{12,34}^{(4,\mathrm{vac})} G_{35}^{(\mathrm{vac})}
  \Gvac_{46} \Lambda_{56,1'2'}}{3456}.
\end{equation}
Once this logarithmically divergent vacuum subdivergence is renormalized
also $\Sigma^{(0,\mathrm{div})}$ is finite since $G^{(r)}$ is falling off
with momentum power $-6$.

For the renormalization of the four-point function we note that the
momentum-space version of (\ref{36}) reads
\begin{equation}
\label{37}
\begin{array}{ll}
\Lambda^{(\mathrm{vac})}(p,q)&=\Gamma^{(4,\mathrm{vac})}(p,q) + \ii
\feynint{l} \Gamma^{(4,\mathrm{vac})}(p,l) [\Gvac(l)]^2
\Lambda^{(\mathrm{vac})}(l,q) \\[1mm]
&=\Gamma^{(4,\mathrm{vac})}(p,q) + \ii
\feynint{l} \Lambda^{(\mathrm{vac})}(p,l)  [\Gvac(l)]^2
\Gamma^{(4,\mathrm{vac})}(l,q)
\end{array} 
\end{equation}
To renormalize this equation, a detailed BPHZ-analysis uses the fact that
due to the 2PI-feature of the $\Phi$ functional
$\Gamma_{12,34}^{(4,\mathrm{vac})}$ is 2PI relative to any cut which
separates the space-time point pairs $(12)$ and $(34)$. Thus there is no
``BPHZ-box'' cutting through this Bethe-Salpeter kernel. Thus we can do the
subtractions at the upper and the lower end of any subdivergence with the
renormalized BS-kernel leading to the \emph{renormalized BS-equation}
\begin{equation}
\label{38}
\begin{array}{ll}
\displaystyle
\Lambda^{(\mathrm{ren})}(p,q) &= \Gamma^{(4,\mathrm{ren})}(p,q) 
\\[2mm]
&+\ii \fint{l}
[\Gamma^{(4,\mathrm{ren})}(p,l)-\Gamma^{(4,\mathrm{vac})}(0,l)]
[\Gvac(l)]^2 \Lambda^{(\mathrm{ren})} \\[2mm]
\displaystyle
& +\ii \fint{l} \Lambda^{(\mathrm{ren})}(0,l) [\Gvac(l)]^2
[\Gamma^{(4,\mathrm{ren})}(l,q) - \Gamma^{(4,\mathrm{ren})}(l,0)].
\end{array}
\end{equation}
The renormalization of $\Gamma^{(4,\mathrm{vac})}$ itself is straight
forward, since it is given by a finite set of vacuum diagrams which can be
renormalized by the same BPHZ-scheme as the perturbative ones.

\begin{figure}
\centering{
\begin{minipage}{0.49\textwidth}
\centering{\includegraphics[width=\textwidth]{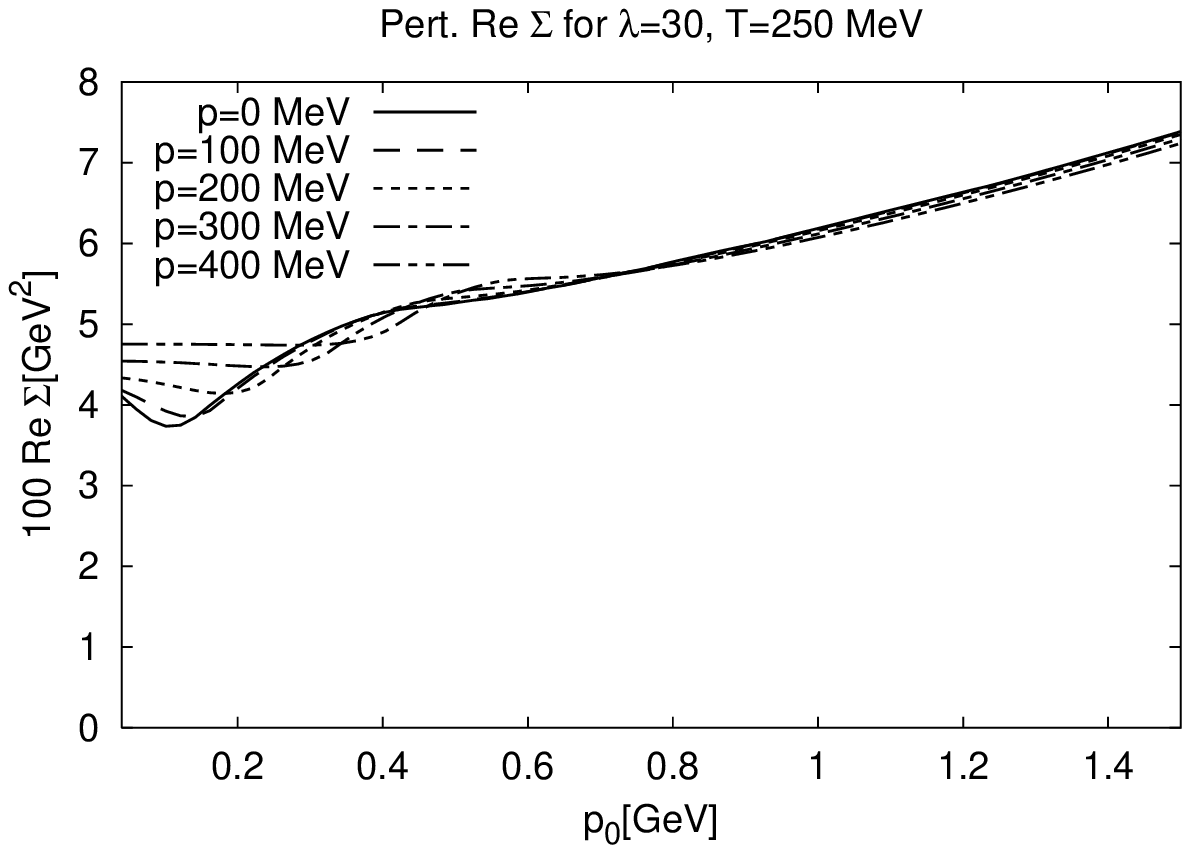}}
\end{minipage}\hspace*{0.2cm}
\begin{minipage}{0.49\textwidth}
\centering{\includegraphics[width=\textwidth]{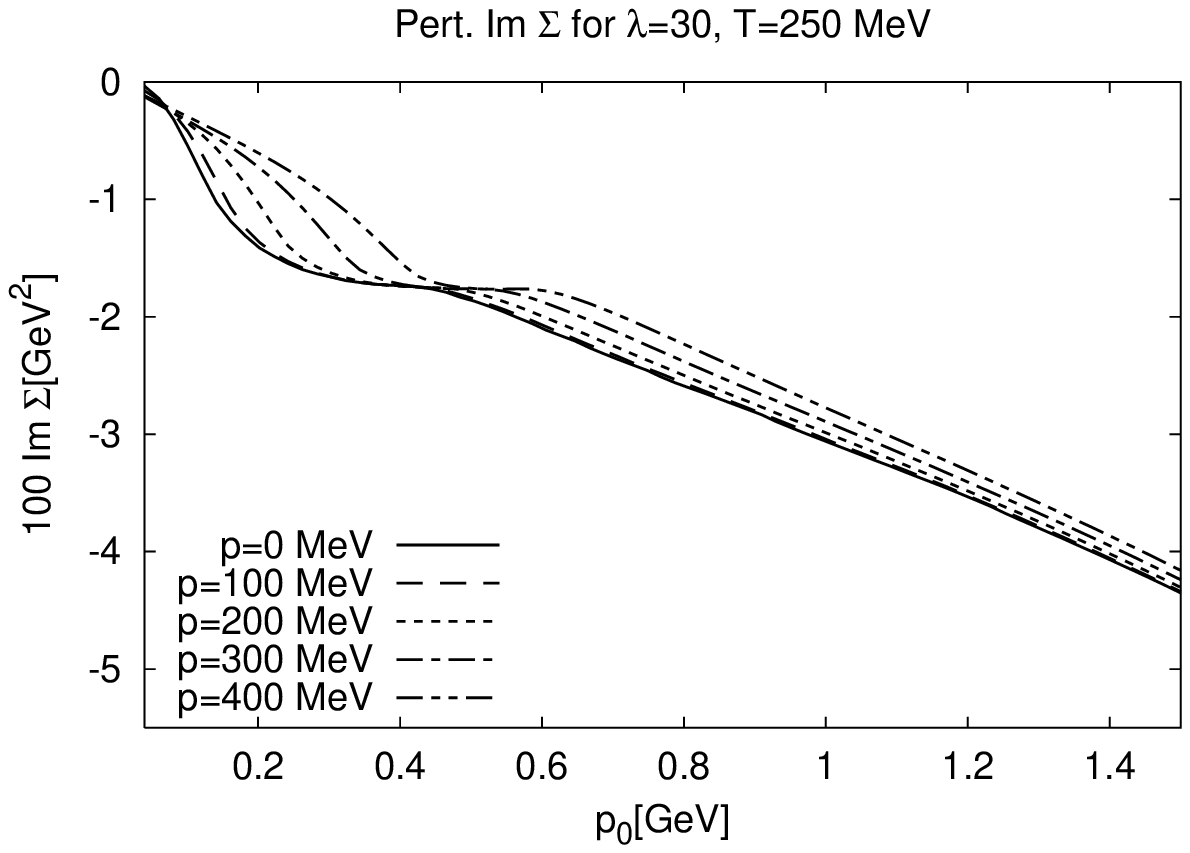}}
\end{minipage}}
\vspace{10mm}
\centering{
\begin{minipage}{0.49\textwidth}
\centering{\includegraphics[width=\textwidth]{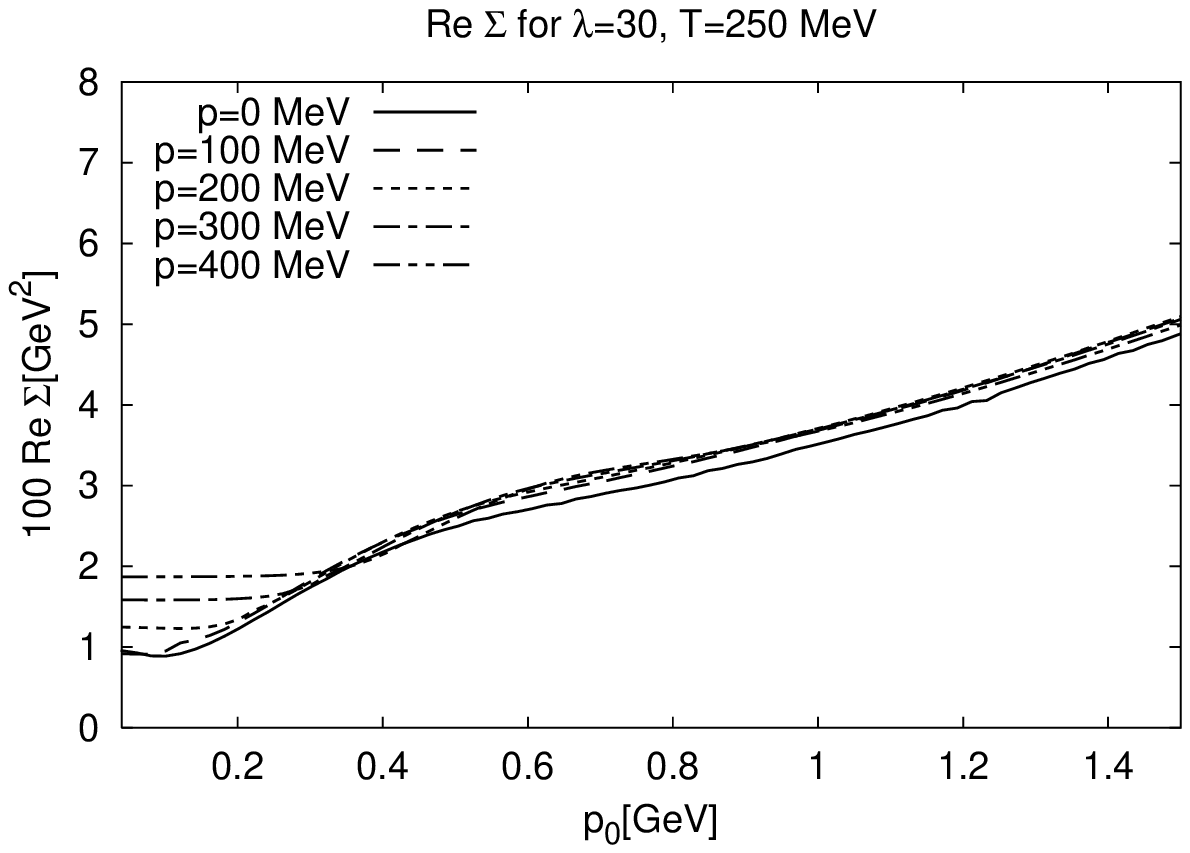}}
\end{minipage}\hspace*{0.2cm}
\begin{minipage}{0.49\textwidth}
  \centering{\includegraphics[width=\textwidth]{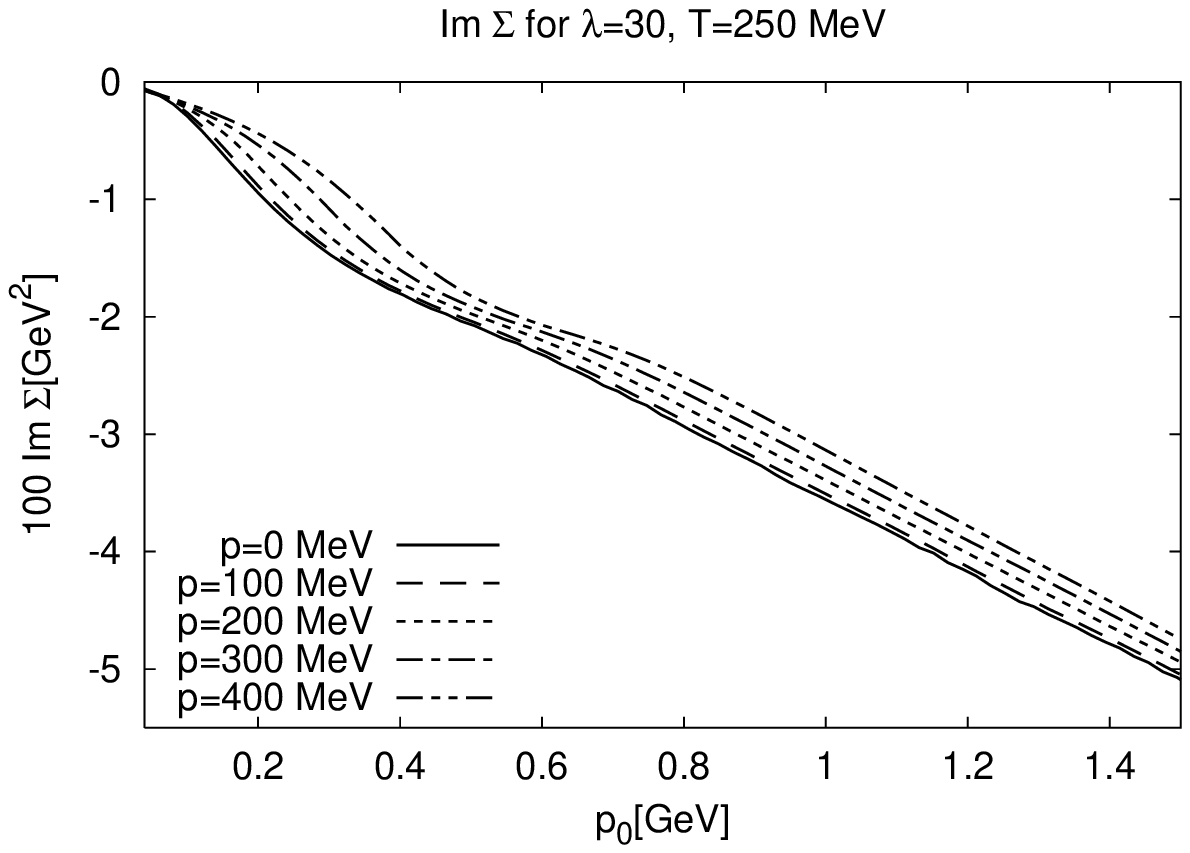}}
\end{minipage}
}
\caption{Real (left) and imaginary part (right) of the perturbative
  (top) and the self-consistent self-energy for $\lambda=30$,
  $m=140\mathrm{MeV}$ and $T=250 \mathrm{MeV}$. Note that the self-energies are
  multiplied with an factor $100$ in these plots!}
\label{fig:3} 
\end{figure}

For the practical calculation of the self-energy part $\Sigma^{(0)}$ we
need only the $\Lambda^{(\mathrm{ren})}(0,q)$. Indeed using Eqs. (\ref{33})
and (\ref{35}) we find
\begin{equation}
\label{39}
\begin{array}{ll}
\displaystyle
\Sigma^{(0)}(p) &= \Sigma^{(0)}(p) - \Sigma^{(0)}(p) + \Sigma^{(0)}(p) \\[1mm]
\displaystyle 
&= \fint{l} [\Gamma^{(4,\mathrm{ren})}(p,l)-\Gamma^{(4,\mathrm{ren})}(0,l)]
G^{(\mathrm{matter})}(l) \\[1mm]
\displaystyle
& + \fint{l} \Lambda^{(\mathrm{ren})}(0,l) G^{(r)}(l)
\end{array}
\end{equation}
An example solution for the equations and the comparison with the
perturbative approximation is shown in Fig. \ref{fig:3}.

While at $T=0$ both, the perturbative and the self-consistent solution,
show the three-particle threshold at $\sqrt{s}=3m$ at finite temperature
the spectral width smoothes out all threshold structures in the
self-consistent solution. The growing high-energy tail is related to the
decay of a virtual particle to three particles. At finite temperature as an
additional effect a low-energy plateau in $\im \Sigma^R$ emerges from
in-medium scattering processes of real particles.

The comparison of the self-consistent solutions with the perturbative
approximation shows counterbalancing effects of self-consistency: The
finite spectral width, contained in the self-consistent propagator leads to
a further broadening of the width and a smoothing of the structure as a
function of energy. This is counterbalanced by the behavior of the real
part of the self-energy which essentially shifts the in-medium mass
upwards. This reduces the available phase space for real processes. With
increasing coupling strength $\lambda$ a nearly linear behavior of $\im
\Sigma^R$ with $p_0$ results, implying a nearly constant damping width,
given by $-\im \Sigma^R/p_0$.

While the tadpole contribution always shifts the mass upwards, at higher
couplings and temperature also the real part of the sunset diagram becomes
significant which can lead to a net down-shift of the mass again. This can
be seen for the parameter set used for the Fig. \ref{fig:3}.

\section{Global symmetries} 
\label{sect:symm}

If the classical action underlying a quantum field theory is invariant
under the linear operation of a Lie group $G$, where the group elements are
independent of the space-time argument one speaks of a \emph{global
  symmetry\index{global symmetry} of the classical action}. Then it can be
shown that for each linearly independent generator $t^a \in \mathcal{L} G$
there exists a conserved charge\index{conserved charge} $Q^{a}$ which in
turn builds a basis of the Lie algebra $\mathcal{L} G$ as a subalgebra of
the canonical Poisson algebra of the fields.

It is a well-known theorem in perturbative quantum field theory that the
same holds true for the quantized theory, i.e., the effective action is
symmetric under the same symmetry group as the classical action provided
that no quantum anomaly destroys the symmetry or a part of it. The
$n$-point Green's functions fulfill certain constraints due to the symmetry
which we shall call the \emph{Ward-Takahashi identities
  (WTIs)}\index{Ward-Takahashi identities} of the symmetry.

For $\Phi$-derivable approximations in general the WTIs of the symmetry are
violated for the two-point and higher vertex functions although the
expectation values of the corresponding Noether currents are exactly
conserved. This can be traced back to a violation of crossing symmetry
within the self-consistent propagator due to partial resummation of the
self-energy insertions by means of the Dyson equation of motion. We shall
further show that the WTIs are fulfilled for vertex functions, which we
shall call \emph{external}. These functions are defined from a
non-perturbative effective action which is uniquely determined by the
underlying $\Phi$-derivable approximation.

We study the symmetry properties of the generating functional (\ref{8})
of a scalar O($N$)-symmetric quantum field theory. An infinitesimal
O($N$)-transformation reads $\delta \phi_1^j = \ii \delta \chi_1^a
{(\tau^a)^j}_{j'} \phi_1^{j'}$. Making use of the assumed invariance of the
classical action under O($N$)-transformations one finds by standard
path-integral analysis\cite{vHK2001-Ren-III} that $\Gamma[\varphi,G]$ is an
O($N$)-scalar functional when $\phi$ is transformed as a vector and $G$ as
a second-rank tensor:
\begin{equation}
\label{46}
\funcint{\funcd{\Gamma}{\phi_1^j} {(\tau^a)^{j}}_{k} \phi_{1}^{k}}{1} +
\funcint{ \funcd{\Gamma}{G_{12}^{jk}} [{(\tau^a)^j}_{j'} G_{12}^{j'k} +
  {(\tau^{a})^k}_{k'} G_{12}^{jk'}]}{12}=0.
\end{equation}
Now only for the exact functional the self-consistent Green's function $G$
is identical with the exact one. Thus in general only for the exact case
the self-energy and higher vertex functions fulfill all WTIs of the usual
1PI action.

Generally for a $\Phi$-derivable approximation this equality of vertex
functions does not hold true any longer. For such approximations in general
the WTIs are violated in loop orders higher than that taken into
account for the approximation of the functional. The reason is that the
solution of the equations of motion are equivalent to a certain partial
resummation of the perturbation series which is not crossing symmetric in
the inner structure of the diagrams.

To recover the crossing symmetry we define an effective action functional
from our approximated 2PI-action as
\begin{equation}
\label{47}
\tilde{\Gamma}[\varphi]= \Gamma[\varphi,\tilde{G}[\varphi]] \mbox{ with }
\left . \funcd{\Gamma[\varphi,G]}{G} \right |_{G=\tilde{G}[\varphi]}=0,
\end{equation}
i.e., for an arbitrarily given mean field $\varphi$ we define the
propagator $\tilde{G}[\varphi]$ as the solution of the Dyson equation of
motion as it is defined by the 2PI-functional. The approximation to the 1PI
functional $\tilde{\Gamma}[\varphi]$ is given by insertion of this
propagator in the 2PI functional. If one would not apply any approximations
$\tilde{\Gamma}$ is identical with the usual 1PI quantum action functional
generating 1PI truncated \emph{proper vertex functions}.
 
The stationary point of this action functional defines the mean field
and propagator of the $\Phi$-derivable approximation since
\begin{equation}
\label{48}
\funcd{\tilde{\Gamma}}{\varphi}=\left ( \funcd{\Gamma[\varphi,G]}{\varphi}
  + \funcint{\funcd{\Gamma[\varphi,G]}{G_{12}^{jk}}
  \funcd{\tilde{G}_{12}^{jk}[\varphi]}{\varphi}}{12}
  \right)_{G=\tilde{G}[\varphi]}.
\end{equation}
From (\ref{47}) this yields
\begin{equation}
\label{49}
\funcd{\tilde{\Gamma}}{\varphi}  =\left (
  \funcd{\Gamma[\varphi,G]}{\varphi} \right)_{G=\tilde{G}[\varphi]}.
\end{equation}
Thus the stationary point $\tilde{\varphi}$ of $\tilde{\Gamma}$ is
identical to the mean field of the $\Phi$-derivable approximation and
together with (\ref{47}) this means that $\tilde{G}[\tilde{\varphi}]$ is
the solution of the Dyson equation from the same $\Phi$-functional.

For a $\Phi$-derivable approximation $\tilde{\Gamma}$ defines a
non-perturbative approximation to this functional and can be used to derive
approximations for the proper vertex functions:
\begin{equation}
\label{50}
(\tilde{\Gamma}^{(n)})_{12\ldots}^{jk\ldots}=\frac{\delta^{n}
  \tilde{\Gamma}[\varphi]}{\delta \varphi_1^j \delta \varphi_2^k \cdots}.
\end{equation}
Especially $\tilde{\Gamma}^{(2)}$ is an approximation for the inverse
propagator, which we call the \emph{external propagator} to be
distinguished from the self-consistent propagator. This external propagator
fulfills the usual WTI but is not identical with the self-consistent or
\emph{internal} propagator.

The vertex functions (\ref{50}) are by definition crossing symmetric. Now
the symmetry property (\ref{46}) by construction holds also true for the
approximated $\Phi$-functional and thus from (\ref{47}) we see that
$\tilde{\Gamma}$ is an O($N$)-scalar functional:
\begin{equation}
\label{51}
\funcint{\funcd{\tilde{\Gamma}[\varphi]}{\varphi_1^j} {(\tau^a)^j}_{j'}
  \varphi_1^{j'}}{1} = 0.
\end{equation}
This symmetry property contains all WTIs for the external vertex functions.
Especially for the external propagator, defined by
\begin{equation}
\label{52}
(G_{\mathrm{ext}}^{-1})_{1j,2k} = \left . \frac{\delta^2
  \tilde{\Gamma}[\varphi]}{\delta \varphi_1^j \delta \varphi_2^k} \right|_{\varphi=\tilde{\varphi}}.
\end{equation}
By taking the functional derivative of (\ref{51}) we obtain
\begin{equation}
\label{53}
\funcint{(G_{\mathrm{ext}}^{-1})_{1j,2k} {(\tau^a)^j}_{j'}
\tilde{\varphi}_{1}^{j'}}{1} =0.
\end{equation} 
For a translationally invariant state the Fourier transform of (\ref{53})
with respect to $(x_1-x_2)$ reads
\begin{equation}
\label{54}
(G_{\mathrm{ext}}^{-1})_{jk}(p=0) {(\tau^a)^j}_{j'} \varphi^{j'} =
-(M^2)_{jk} {(\tau^a)^j}_{j'} \tilde{\varphi}^{j'} = 0.
\end{equation}
It is clear that in this situation $\tilde{\varphi}$ is a constant due to
translation invariance. If it is not $0$ the symmetry is spontaneously
broken. Since $(M^2)_{jk}$ is the mass matrix of the particles described as
the excitations of the field around the mean field $\tilde{\varphi}$ Eq.
(\ref{54}) tells us that the $N-1$ fields perpendicular to the direction
given by the solution $\tilde{\varphi}$ are massless, the
\emph{Nambu-Goldstone bosons}\index{Nambu-Goldstone bosons} of the
symmetry.

To calculate the external propagator explicitly we apply (\ref{47}) in
(\ref{52}) to obtain
\begin{equation}
\label{55}
(G_{\mathrm{ext}}^{-1})_{1j,2k} = 
\left [ \frac{\delta^2 \Gamma[\varphi,G]}{\delta \varphi_1^j \delta
  \varphi_2^k}  + \funcint{\frac{\delta^2 \Gamma[\varphi,G]}{\varphi_1^j
  \delta G_{3'4'}^{j'k'}}
  \funcd{\tilde{G}_{3'4'}^{j'k'}}{\varphi_2^k}}{3'4'} \right
  ]_{\varphi=\tilde{\varphi}, \; G=\tilde{G}[\tilde{\varphi}]}.
\end{equation}
Taking the derivative of the identity
\begin{equation}
\label{56}
\funcint{(\tilde{G}^{-1})_{1j,2'k'}
  \tilde{G}_{2'2}^{k'k}}{2'}=\delta_{12}^{(d)} \delta_{j}^{k}
\end{equation}
with respect to the field we get
\begin{equation}
\label{57}
\funcint{\left [\funcd{\tilde{G}^{-1}}{\varphi_3^l} \tilde{G}_{2'2}^{k'k}
    + (\tilde{G}^{-1})_{1j,2'k'} \funcd{G_{2'2}^{k'k}}{\varphi_3^l}
    \right ]}{1'2'}=0.
\end{equation}
With help of (\ref{12}) the three-point function
\begin{equation}
\label{58}
\Lambda_{1j,2k;3l}^{(3)}=\funcd{\tilde{G}^{-1}_{1j,2k}}{\varphi_3^l}
\end{equation}
can be expressed as the solution of the BS-equation
\begin{equation}
\label{59a}
\Lambda_{1j,2k;3l}^{(3)}=\Gamma_{1j,2k;3l}^{(3)} - \ii
\funcint{\Gamma_{1j,2k;3'l'4'm'}^{(4)} \tilde{G}_{3'3''}^{l'l''}
  \tilde{G}_{4' 4''}^{m' m''} \Lambda_{3''l'',4''m'';3l}^{(3)}}{3'4'3''4''},
\end{equation}
which in terms of diagrams can be depicted as
\begin{equation}
\label{59}
\ii \Lambda^{(3)}=\parbox{9.12cm}{\includegraphics[scale=0.8]{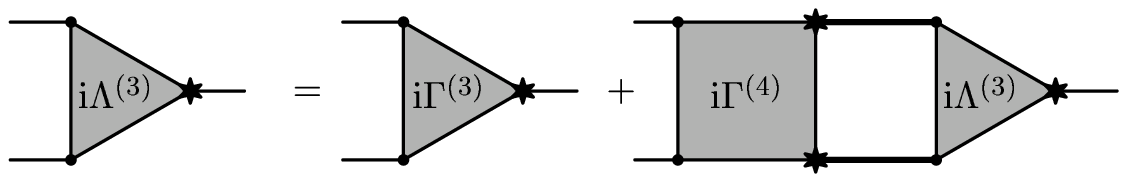}},
\end{equation}
while its kernels are determined by the $\Phi$-functional:
\begin{equation}
\label{60}
\begin{array}{ll}
\displaystyle \Gamma_{1j,2k;3l}^{(3)} &= \displaystyle \left [ \frac{\delta^3 S[\varphi]}{\delta
    \varphi_1^j  \delta \varphi_2^k \delta \varphi_3^l} - 2 \ii
    \frac{\delta^2 \Phi[\varphi,G]}{\delta G_{12}^{jk} \delta \varphi_3^l}
    \right ]_{G=\tilde{G}[\tilde{\varphi}], \; \varphi=\tilde{\varphi}} \\[4mm]
\displaystyle \Gamma_{1j,2k;3l,4m}^{(4)} &=\displaystyle -2 \left [\frac{\delta^2
    \Phi[\varphi,G]}{\delta G_{12}^{jk} \delta G_{34}^{lm}}
    \right]_{G=\tilde{G}[\tilde{\varphi}], \; \varphi=\tilde{\varphi}} 
\end{array}
\end{equation}
With these definitions using (\ref{57} and (\ref{58}) the external
propagator (\ref{55}) can be written as
\begin{equation}
\label{62a}
\begin{array}{ll}
\displaystyle (\Sigma_{\mathrm{ext}})_{1j,2k} = & \displaystyle - \Bigg
[\frac{\ii}{2} 
\funcint{\frac{\delta^4 S[\varphi]}{\delta \varphi_1^j \delta \varphi_2^k
    \delta \varphi_1^{j'} \delta \varphi_{2'}^{k'}} G_{1'2'}^{j'k'}}{1'2'} 
  \displaystyle  +
  \frac{\delta \Phi[\varphi,G]}{\delta \varphi_1^j \delta \varphi_2^k}
  \displaystyle \\ & \displaystyle -
  \frac{\ii}{2} \funcint{\Gamma_{3'j'4'k';1j}^{(3)} G_{3'3''}^{j'j''}
    G_{4'4''}^{k'k''} \Lambda_{3''j'',4''k'';2k}}{3'4'3''4''}
  \Bigg ].
\end{array}
\end{equation}
In graphical terms this equation looks as follows
\begin{equation}
\label{62}
\raisebox{-3mm}{\parbox{10.0cm}{\includegraphics[scale=0.8]{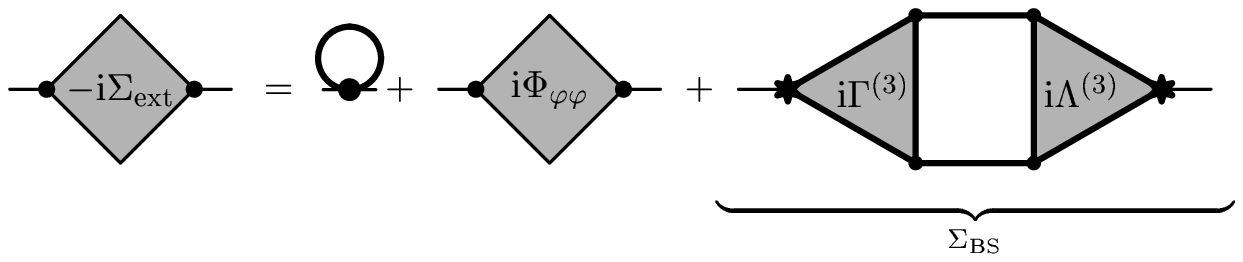}}}
\end{equation}
It is clear that both, the BS-equation (\ref{59a}) and Eq. (\ref{62a}),
have to be renormalized. This is done again with help of the above
explained BPHZ techniques. Again all counterterms turn out to be
independent of the temperature and consistent with those needed to
renormalize the underlying self-consistent equations of motion (for more
details of this renormalization see\cite{vHK2001-Ren-III}).

Looking at the diagrams (\ref{59}) and (\ref{62}) it turns out that the
BS-equation (\ref{59a}) provides exactly the resummation of the channels
missing inside the self-consistent approximation to the self-energy. This
restores both the intrinsic \emph{crossing symmetry} and the underlying
O($N$)-WTI for the \emph{external self-energy}.

\begin{figure}
\begin{minipage}{0.45\textwidth}
\centerline{\includegraphics[height=50mm]{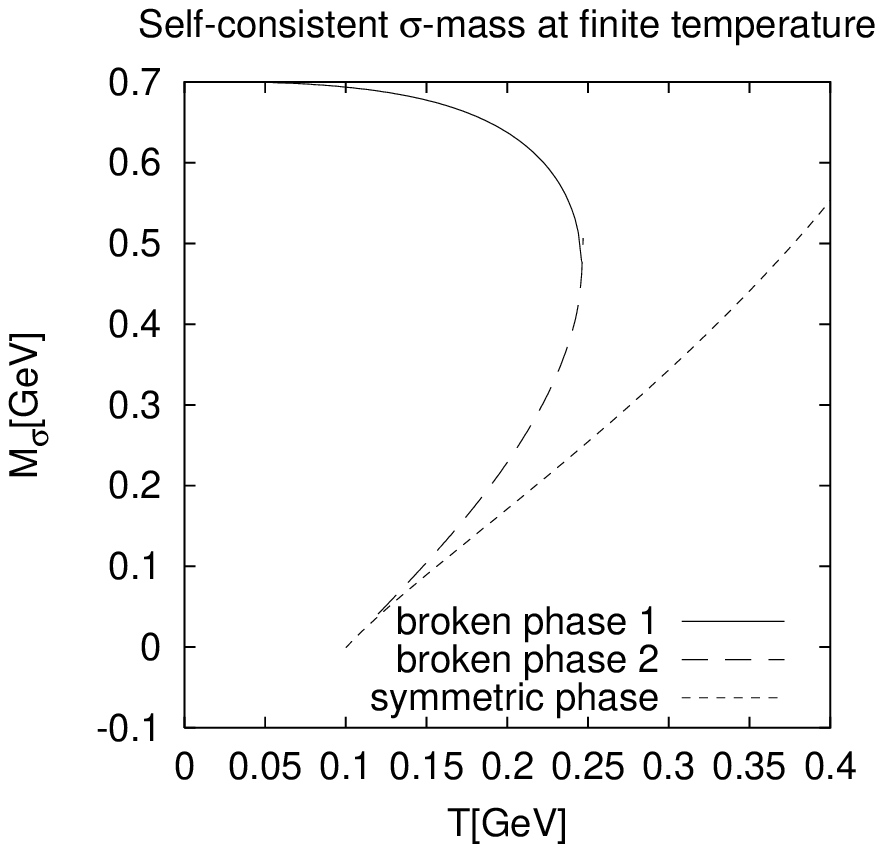}}
\end{minipage}
\begin{minipage}{0.45\textwidth}
\centerline{\includegraphics[height=50mm]{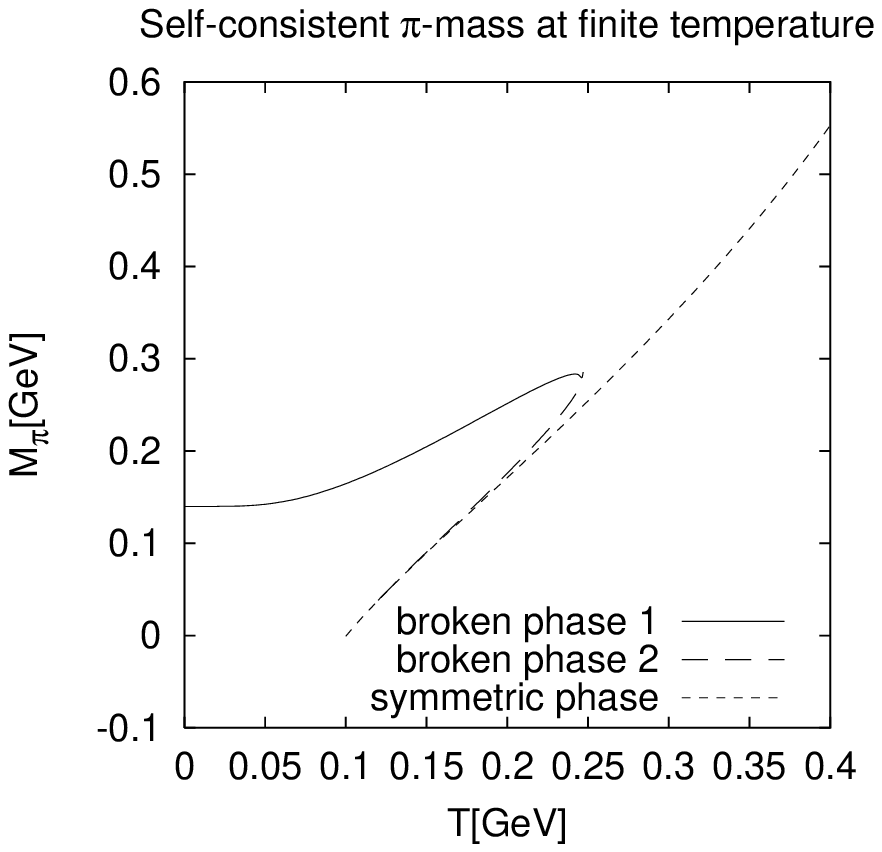}}
\end{minipage}
\caption{The solutions for the Hartree approximation}
\label{fig:4}
\end{figure}

As an example we show results for the lowest order approximation which
is the Hartree approximation for the self-consistent self-energy, leading
to a constant effective mass shown in Fig. \ref{fig:4}.
Clearly Goldstone's theorem\index{Goldtone's theorem} is violated, since
the pion mass is different from $0$ in both broken phases.
\begin{figure}
\begin{minipage}{0.4\textwidth}
\centerline{\includegraphics[width=\textwidth]{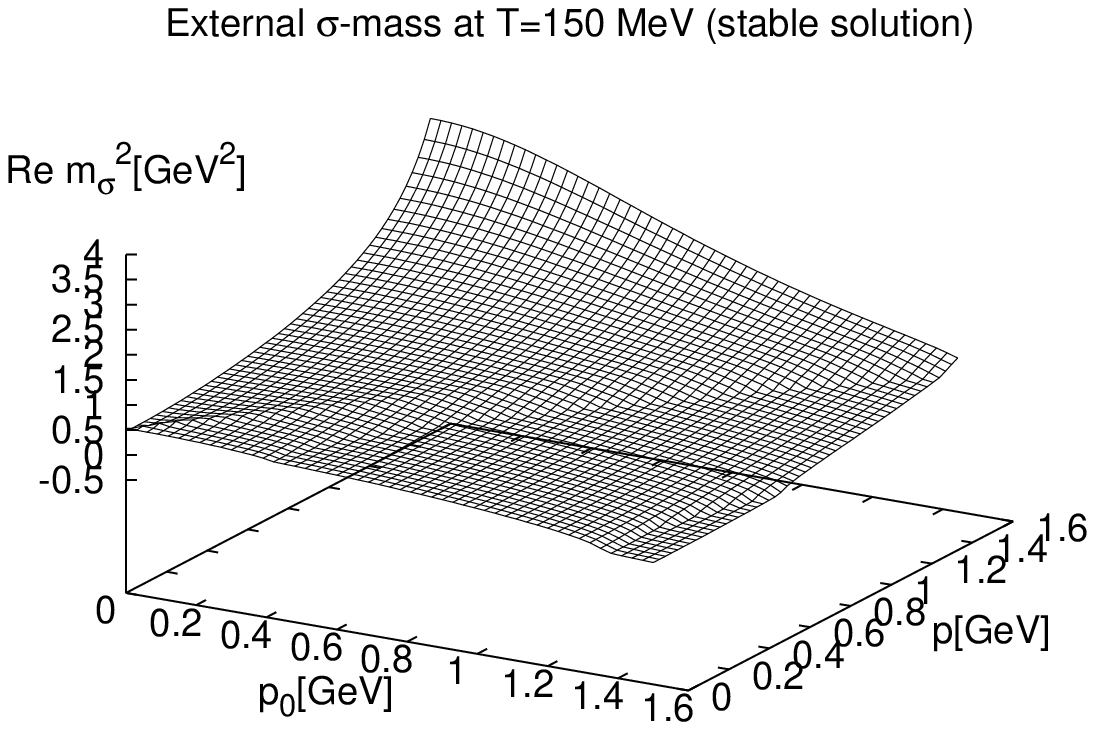}}
\vspace*{5mm}
\end{minipage}
\begin{minipage}{0.4\textwidth}
\centerline{\includegraphics[width=\textwidth]{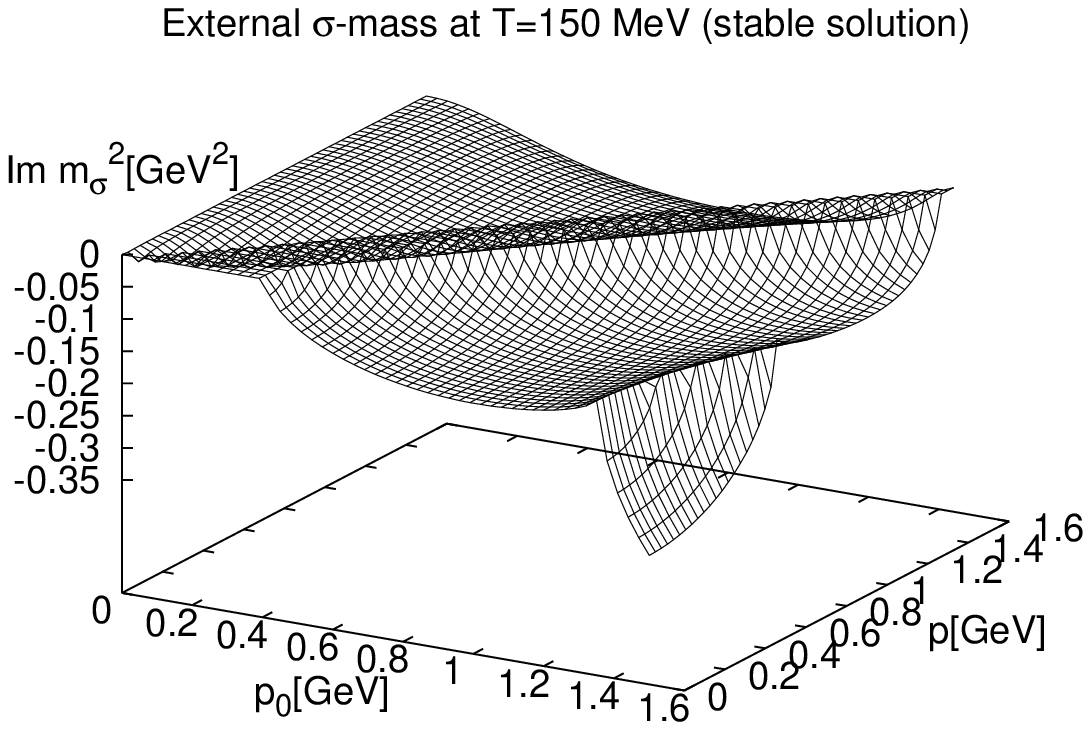}}
\vspace*{5mm}
\end{minipage}\hfill
\begin{minipage}{0.4\textwidth}
\centerline{\includegraphics[width=\textwidth]{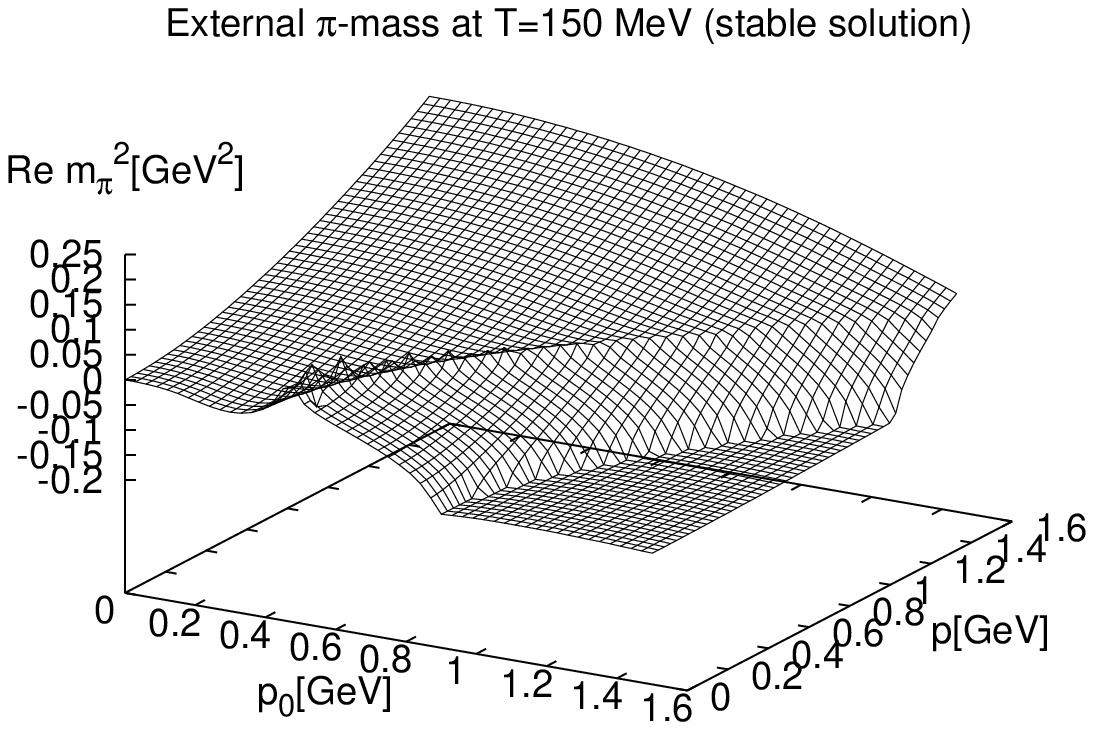}}
\end{minipage}\hfill
\begin{minipage}{0.4\textwidth}
\centerline{\includegraphics[width=\textwidth]{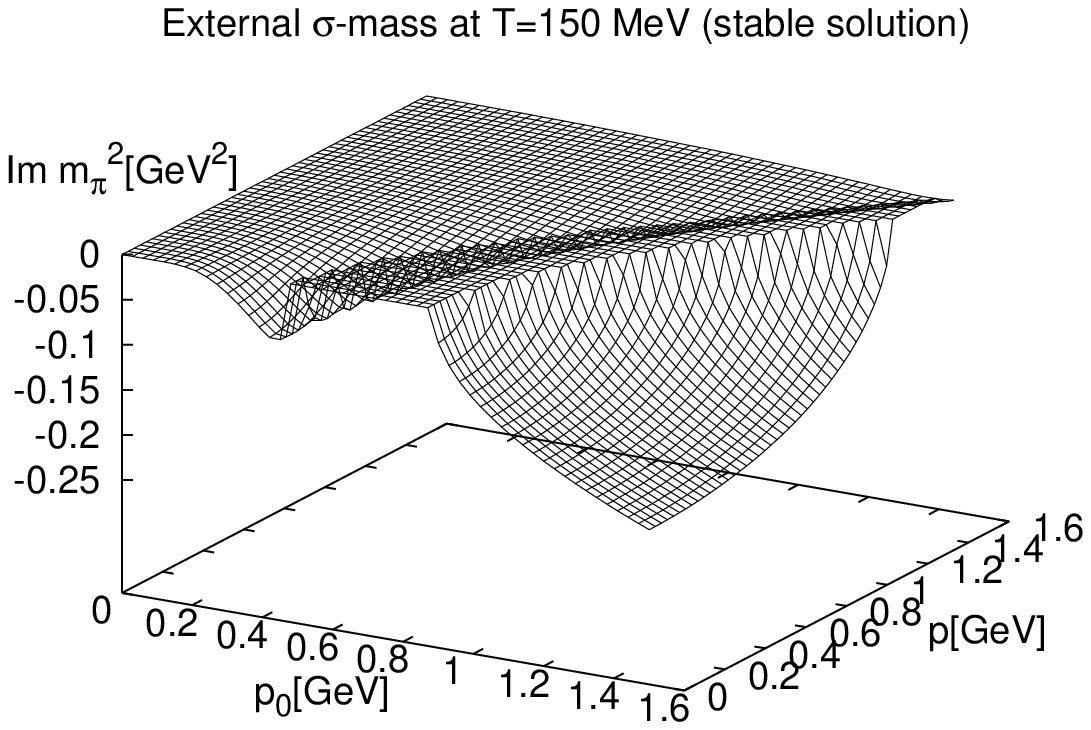}}
\end{minipage}
\caption{The effective external masses at a temperature of $150
  \mathrm{MeV}$. The effective external $\pi$-mass indeed vanishes at
  $p_{0}=\vec{p}=0$ as predicted from Goldstone's theorem. The spectral
  function of the $\sigma$-meson shows that at high temperatures its
  strength becomes more peaked and the maximum shifted to lower momenta
  than at $T=0$.}
\label{fig:5}
\end{figure}
In this case the external self-energy is obtained by a bubble
resummation\index{bubble resummation}, corresponding to a \emph{Random
  phase approximation (RPA)}\index{Random phase approximation}. Since at
the same time it is the second functional derivative of the effective
1PI-action functional it provides a \emph{stability criterion} for the
Hartree solution, which is of course only a stable solution, i.e., a
minimum of the effective potential, if the mass matrix is positive
semidefinite. It turns out that the solution, denoted as ``broken phase 2''
in Fig. \ref{fig:4} is \emph{unstable}. In Fig. \ref{fig:5} the effective
mass obtained from the \emph{external propagator}, is shown. Indeed
Goldstone's theorem is fulfilled, as to be expected from our analysis. From
another point of view this result was also obtained by Aouissat and
Belacem\cite{aou99}.

\section{Conclusions and outlook}

We have shown that any self-consistent Dyson resummation can be
renormalized with counter terms that are independent of temperature
provided it is realized as a $\Phi$-derivable approximation. The proof is
based on the applicability of Weinberg's power counting theorem and the
BPHZ subtraction scheme. These techniques provide the possibility to
extract both, the explicite and hidden divergent vacuum subdiagrams, and to
subtract the divergences leading to a coupled set of finite equations of
motion for the mean fields and the Green's functions.

In the second part a detailed analysis of symmetry properties has been
given. In general the $\Phi$-derivable approximation schemes violate
Ward-Takahashi identities for two-point and higher Green's functions since
intrinsically crossing symmetry is violated at orders of the expansion
parameter higher than that taken into account for the 2PI functional. Also
the recently observed deviation from the correct renormalization group flow
of the coupling constant can be traced back to this incomplete resummation
provided by the self-consistent equations of motion\cite{bp01}.

It was further shown that the symmetries are recovered by defining a
non-perturbative approximation to the effective quantum action by the
stationary point of the 2PI action functional with respect to the Green's
function at given mean fields. The calculation of the self-energy defined
with help of this 1PI functional needs the solution of a Bethe-Salpeter
equation keeping track of the channels that are missing intrinsically in
the $\Phi$-derivable self-consistent resummation for the self-consistent
propagator. In this way not only crossing symmetry but also the
Ward-Takahashi identities for the proper vertex functions related to
linearly realized global symmetries are recovered.

The problems related with the missing crossing symmetry and symmetry
violations become more serious in the case of local gauge symmetries. In
general then the $\Phi$-derivable approximations violate important features
like unitarity, causality and positive definiteness of the probability
measure. Here only partial solutions of these problems are known, for
instance making use of hard thermal loop expansion
schemes\cite{pesh00,ianc00}. A systematic analysis of the violation of
gauge invariance has been given recently by Arrizabalaga and
Smit\cite{as2002}. It was shown by Denner and Dittmaier\cite{den96} that
there exist non-symmetry breaking Dyson resummation schemes within the
background field gauge formulation. At the time an investigation whether or
not this scheme is applicable also in the context of the $\Phi$-derivable
formalism is under way. Nevertheless the definition of the self-consistent
propagator is in this case questionable because of the artificial
excitation of unphysical gauge field degrees of freedom by the violation of
crossing symmetry inside the self-consistent diagrams. A first way out of
this problem by a projection formalism is given by us in\cite{vHK2001}.

For the study of non-equilibrium situation the $\Phi$-functional becomes
the only systematic approximation scheme which obeys conservation laws.
Recently numerical studies of the equations of motion for out of
equilibrium were undertaken (see \cite{berges00a,berges01a,berges01b,ahr02}
and citations therein).

The $\Phi$-derivable approximations are also a starting point for the
derivation of transport equations from quantum field theory, ensuring
consistency conditions like detailed ballance, Boltzmann's H-theorem and
conservation laws.  Especially it becomes important for finding consistent
transport equations for particles or resonances with a finite mass
width\cite{kv97,ikv99-2,ivvH00,KIV01}.

\vspace*{0.3mm} \noindent \textbf{Acknowledgment}

We are grateful to D. Ahrensmeier, R. Baier, J. Berges, J. P.  Blaizot, P.
Danielewicz, B.  Friman, E. Iancu, Yu. Ivanov, M. Lutz, L. McLerran,
E. Mottola, R. Pisarski, and D.  Voskresensky for fruitful discussions
and suggestions at various stages of this work.

\begin{flushleft}

\end{flushleft}

\printindex

\end{document}